\documentclass[twocolumn,pra,floatfix,showpacs,superscriptaddress]{revtex4}
\usepackage{amssymb}
\usepackage{amsmath}
\usepackage{graphicx}
\usepackage{float}

\newcommand\be{\begin{eqnarray}}
\newcommand\ee{\end{eqnarray}}

\begin{document}

\title{Propagation of Light in an Ensemble of "3+1"-Level Atoms}
\author{Yong Li}
\affiliation{Institute of Theoretical Physics, Chinese Academy of Sciences, Beijing,
100080, China}
\affiliation{Interdisciplinary Center of Theoretical Studies, The Chinese Academy of
Sciences, Beijing, 100080, China}
\affiliation{Department of Physics, Hunan Normal University, Changsha, 410081, China}
\author{C. P. Sun}
\email{suncp@itp.ac.cn} \homepage{http://www.itp.ac.cn/~suncp }
\affiliation{Institute of Theoretical Physics, Chinese Academy of Sciences,
Beijing, 100080, China} 

\begin{abstract}
We study the propagation of a quantum probe light in an ensemble of
"3+1"-level atoms when the atoms are coupled to two other classical control
fields. First we calculate the dispersion properties, such as susceptibility
and group velocity, of the probe light within such an atomic medium under
the case of three-photon resonance via the dynamical algebra method of
collective atomic excitations. Then we calculate the dispersion of the probe
light not only under the case that two classical control fields have the
same detunings to the relative atomic transitions but also under the case
that they have the different detunings. Our results show in both cases the
phenomenon of electromagnetically induced transparency can accur. Especially
use the second case, we can find two transparency windows for the probe
light.
\end{abstract}

\maketitle

\section{Introduction}

The coherent interaction of atoms with optical fields has attracted much
attention in studies of contemporary coherent and nonlinear optics
\cite{scully2,harris,fleischhauer99}. One of the most interesting effects is
electromagnetically induced transparency (EIT) \cite{EIT}. In an EIT system,
the atoms are optically pumped into a {\it so-called} dark state which is
decoupled from the original optical fields. Such an atomic medium possesses
special optical properties such as cancellation of resonant absorption and
slow group velocity of the reference probe light field.

Generally, a conventional EIT system consists of a vapor cell with 3-level $
\Lambda $-type (or $V$-type and cascade type) atoms resonantly coupled to
two classical fields \cite{Marangos98JMO,Boon99PRA}, which are called as the
control and probe light field respectively. Now, many advanced studies have
been done in the field of EIT. On the one hand, people find the EIT
phenomenon can appears not only in the case of exactly one-photon resonance
but also in the case of two-photon resonance
\cite{Deng01,Deng02,li-sun-prar}. And people has done many studies on the
EIT system involving 4-level (or multi-level) atoms
\cite{Un,duan-science,Gu03PRA}. On the other hand, the quantum probe light
is introduced to replace the weak classical probe light field in the EIT
system \cite{Lukin00-ent,Fl00-pol}. Recently, for example, an ensemble of
$\Lambda $-type atoms, where the weak classical probe light is replace by a
quantum probe light to form an EIT system, has been proposed
\cite{Lukin00-ent,Fl00-pol,Fl00-OptCom,Lukin-RMP} as a candidate for
practical quantum memory to store and transfer the quantum information
contained in photonic states by the collective atomic excitations. Some
experiments \cite{liu,group} have also already demonstrated the central
principle of this technique, namely, the reduction of the group velocity and
resonant absorbtion of light.

Moreover, based on these previous work, we have studied a system with
quasi-spin wave collective excitations of many $\Lambda $-type atoms fixed
in "atomic crystal". A hidden dynamical symmetry is discovered in such a
system, and it is considered as a candidate for a robust quantum memory \cite%
{Sun-prl}. It is observed that in certain cases \cite{Sun-quant-ph} the
quantum state can be retrieved up to a non-Abelian Berry phase, i.e., a
non-Abelian holonomy \cite{BPF,ZW,Zana,duan-science,Ekert}, in such a $%
\Lambda $-type atomic system or a similar "3+1"-level atomic ensemble system
\cite{li-zhang}. This observation extends the concept of quantum information
storage and means that the stored state can be decoded in a purely geometric
way in such a case.

The above work about "3+1"-level atomic ensemble only considers the transfer
(or quantum storage) of photonic state within the atomic ensemble. In order
to achieve a complete process of photonic quantum state storage, generally
the probe light should has a slow group velocity in order to make sure it
being within the atomic ensemble during the time of state transfer. In this
work, we shall calculate the dispersion properties of the quantum
probe light field in a "3+1"-level atomic system given in Ref. \cite%
{li-zhang} by means of the novel algebraic dynamics method of atomic
collective excitation shown in Refs. \cite{Sun-prl,li-sun-prar,jin-prb}. By
studying the susceptibility and group velocity of the quantum light, we will
show in what cases this system appears as an EIT one and investigate how the
group velocity depends on the detuning of the control and probe fields.

\section{The Model}

The model we considered consists of $N$ identical "3+1"-level atoms \cite%
{li-zhang,Un,duan-science}, where each atom is coupled to two single-mode
classical control fields and a quantum probe field as shown in Fig. \ref%
{fig1}. The atomic levels are labelled as the ground state $|b\rangle $, the
excited state $|a\rangle $, and the meta-stable states $|k\rangle $ $(k=1,2)$%
. The atomic transition $|a\rangle \rightarrow |b\rangle $, with energy
level difference $\omega _{ab}$=$\omega _{a}-\omega _{b}$, is coupled to the
probe field of frequency $\omega $ $(=\omega _{ab}-\Delta _{p})$ with the
coupling coefficient $g$; and the atomic transition $|a\rangle \rightarrow
|k\rangle $ $(k=1,2)$, with energy level difference $\omega _{ak}$, is
driven by the classical control field of frequency $\nu _{k}$ $(=\omega
_{ak}-\Delta _{k})$ with Rabi-frequency $\Omega _{k}(t)$.

\begin{figure}[h]
\hspace{10pt}\includegraphics[width=5cm,height=5.4cm]{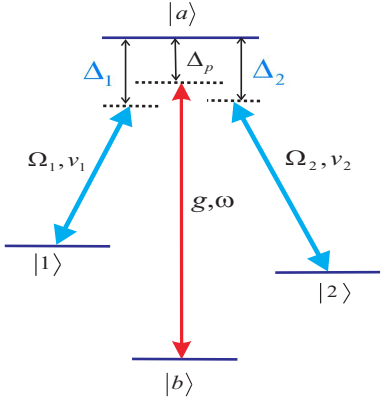}
\caption{"3+1"-type four-level atoms interacting with a quantum probe field
(with coupling constant $g$, frequency $\protect\omega $, and the detuning $%
\Delta _{p}$) and two classic control fields (with frequency $\protect\nu %
_{k}$ , coupling Rabi frequency $\Omega _{k}$, and the detuning $\Delta _{k}=%
\protect\omega _{ak}-\protect\nu _{k}$, $k=1,2$).}
\label{fig1}
\end{figure}

Under the rotating wave approximation we can write the interaction
Hamiltonian in the interaction picture as (let $\hbar =1$) \cite{li-zhang}
\begin{eqnarray}
H_{I} &=&\Delta _{p}S+g\sqrt{N}aA^{\dagger }+\Omega _{1}\exp [i(\Delta
_{1}-\Delta _{p})t]T_{+}^{(1)}  \notag \\
&&+\Omega _{2}\exp [i(\Delta _{2}-\Delta _{p})t]T_{+}^{(2)}+h.c.,
\label{hi01}
\end{eqnarray}%
where
\begin{eqnarray}
S &=&\sum_{j=1}^{N}\sigma _{aa}^{(j)},\text{ }A=\frac{1}{\sqrt{N}}%
\sum_{j=1}^{N}\sigma _{ba}^{(j)},  \notag \\
T_{-}^{(k)} &=&\sum_{j=1}^{N}\sigma _{ka}^{(j)},\text{ }%
T_{+}^{(k)}=(T_{-}^{(k)})^{\dagger },\text{ }(k=1,2)  \label{coll-ops}
\end{eqnarray}%
are symmetrized collective atomic operators. Here $\sigma _{\mu \nu
}^{(j)}=|\mu \rangle _{jj}\langle \nu |$ denotes the flip operator of the $j$%
-th atom from state $|\nu \rangle _{j}$ to $|\mu \rangle _{j}$ $(\mu ,\nu
=a,b,1,2)$; $a^{\dagger }$ and $a$ the creation and annihilation operators
of quantum probe field respectively. The coupling coefficients $g$ and $%
\Omega _{1,2}$ are real and assumed to be identical for all the atoms in the
ensemble.

Let us recall the dynamical symmetry as discovered in Ref. \cite{li-zhang}
in the large $N$ limit and low excitation regime of the atomic ensemble
where most of $N$ atoms stay in the ground state $|b\rangle $. It is obvious
that $T_{-}^{(k)}$ and $T_{+}^{(k)}$ $(k=1,2)$ generate two mutually
commuting $SU(2)$ subalgebras of $SU(3)$ \cite{jin-prb}. To form a closed
algebra containing $SU(3)$ and $\{A,A^{\dagger }\}$ appeared in Hamiltonian (%
\ref{hi01}), two additional collective operators
\begin{equation}
C_{k}=\frac{1}{\sqrt{N}}\sum_{j=1}^{N}\sigma _{bk}^{(j)},\ \ (k=1,2)
\label{C_k}
\end{equation}%
along with their hermitian conjugates are introduced. These operators have
the non-vanishing commutation relations
\begin{eqnarray}
&&C_{k}=[A,T_{+}^{(k)}],\ \ [C_{k},T_{-}^{(k)}]=A,\ \ (k=1,2);  \notag \\
&&[A,A^{\dagger }]=[C_{1},C_{1}^{\dagger }]=[C_{2},C_{2}^{\dagger }]=1.
\label{commuta22}
\end{eqnarray}%
As a special case of quasi-spin wave excitation with zero varying phases,
the above three mode symmetrized excitations defined by $A$ and $C_{1,2}$
behave as three independent bosons.

\section{The susceptibility of quantum probe light field}

Now we will investigate the probe field group velocity from the
time-dependent Hamiltonian (\ref{hi01}). By means of the above dynamic
algebra and commutation relation (\ref{commuta22}), we can write down the
Heisenberg equations of operators $A$ and $C_{1,2}$ as%
\begin{eqnarray}
\dot{A} &=&-(\Gamma _{A}+i\Delta _{p})A-ig\sqrt{N}a  \notag \\
&&-ie^{i(\Delta _{c}-\Delta _{p})t}(\Omega _{1}C_{1}+\Omega
_{2}C_{2})+f_{A}(t), \\
\dot{C}_{1} &=&-\Gamma _{1}C_{1}-ie^{i(\Delta _{p}-\Delta _{c})t}\Omega
_{1}A+f_{1}(t), \\
\dot{C}_{2} &=&-\Gamma _{2}C_{2}-ie^{i(\Delta _{p}-\Delta _{c})t}\Omega
_{2}A+f_{2}(t).
\end{eqnarray}%
Here, we have phenomenologically introduced the decay rates $\Gamma _{{1,2}}$
and $\Gamma _{A}$ of the states $|1\rangle $, $|2\rangle $ and $|a\rangle $,
and $f_{A}(t)$ and $f_{1,2}(t)$ are the relative quantum fluctuation of
operators with $\left\langle f_{\alpha }(t)f_{\alpha }(t^{\prime
})\right\rangle \neq 0$, but $\left\langle f_{\alpha }(t)\right\rangle =0$, $%
(\alpha =A,1,2)$.

To find the steady state solution for the above motion equations of atomic
coherent excitation, it is convenient to remove the fast time-changing
factors by making the transformation $C_{j}=\tilde{C}_{j}\exp [i(\Delta
_{p}-\Delta _{j})t]$ ($j=1,2$). So the transformed equations are given as
\begin{eqnarray}
\dot{A} &=&-(\Gamma _{A}+i\Delta _{p})A-ig\sqrt{N}a  \notag \\
&&-i(\Omega _{1}\tilde{C}_{1}+\Omega _{2}\tilde{C}_{2})+{f}_{A}(t), \\
\dot{\tilde{C}}_{1} &=&-\Gamma _{1}\tilde{C}_{1}-i(\Delta _{p}-\Delta _{1})%
\tilde{C}_{1}-i\Omega _{1}A+\tilde{f}_{1}(t), \\
\dot{\tilde{C}}_{2} &=&-\Gamma _{2}\tilde{C}_{2}-i(\Delta _{p}-\Delta _{2})%
\tilde{C}_{2}-i\Omega _{2}A+\tilde{f}_{2}(t).
\end{eqnarray}%
As shown in Ref. \cite{li-sun-prar}, in the steady state approach and taking
the mean expressions of the above equations, we can obtain
\begin{equation}
ig\sqrt{N}\left\langle a\right\rangle =-F(\Delta _{p})\left\langle
A\right\rangle ,  \label{mean a A}
\end{equation}%
where%
\begin{eqnarray}
F(\Delta _{p}) &=&(\Gamma _{A}+i\Delta _{p})+\frac{\Omega _{1}^{2}}{\Gamma
_{1}+i(\Delta _{p}-\Delta _{1})}  \notag \\
&&+\frac{\Omega _{2}^{2}}{\Gamma _{2}+i(\Delta _{p}-\Delta _{2})}.
\end{eqnarray}%
It is noticed that the single-mode probe quantum light is described by
\begin{equation}
E(t)=\varepsilon e^{-i\omega t}+h.c.\equiv \sqrt{\frac{\omega }{2V\epsilon
_{0}}}ae^{-i\omega t}+h.c.,  \label{E a}
\end{equation}%
where $V$ is the effective mode volume and for simplicity is assumed to be
equal to the volume of the atomic ensemble. While its corresponding
polarization is
\begin{equation}
\left\langle P\right\rangle =\left\langle p\right\rangle e^{-i\omega
t}+h.c.\equiv \epsilon _{0}\chi \left\langle \varepsilon \right\rangle
e^{-i\omega t}+h.c.,  \label{p chi}
\end{equation}%
where $\chi =\left\langle p\right\rangle /(\left\langle \varepsilon
\right\rangle \epsilon _{0})$ is the susceptibility. In terms of the average
of the exciton operators $A$, the average polarization can be expressed as%
\begin{equation}
\left\langle p\right\rangle =\mu \left\langle \sum_{j=1}^{N}\sigma
_{ba}^{(j)}\right\rangle /V=\frac{\mu \sqrt{N}}{V}\left\langle
A\right\rangle ,  \label{mean p A}
\end{equation}%
where $\mu $ is the dipole moment between state $|a\rangle $ and $|b\rangle $%
. It is also noted that the coupling coefficient $g$ can be written as
\begin{equation}
g=-\mu \sqrt{\frac{\omega }{2V\epsilon _{0}}}.  \label{g}
\end{equation}%
According to the Eqs. (\ref{mean a A})-(\ref{g}), the susceptibility can be
obtained as
\begin{equation}
\chi =\frac{2ig^{2}N}{\omega F(\Delta _{p})}.  \label{chi-f}
\end{equation}%
The real and imaginary parts $\chi _{1}$ and $\chi _{2}$ of this complex
susceptibility $\chi =\chi _{1}+i\chi _{2}$ are related to dispersion and
absorption of quantum probe light field, respectively. Here the analytical
solution of $\chi _{1}$ and $\chi _{2}$ is a little complicated to express,
and we only give the numerical solution.

\begin{figure}[h]
\begin{center}
\includegraphics[width=8cm,height=7cm]{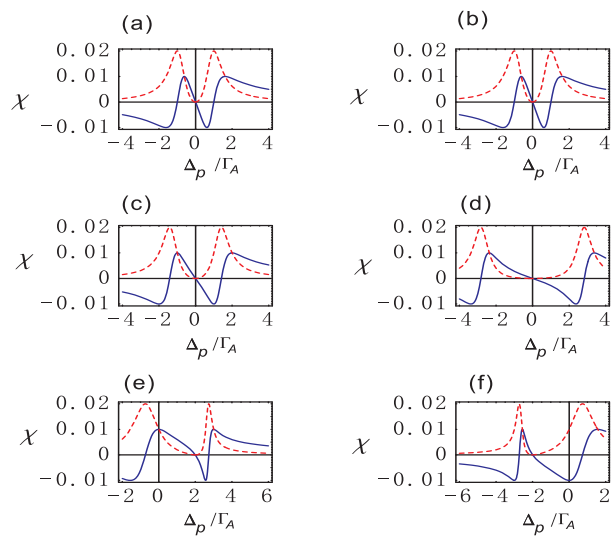}
\end{center}
\caption{Real part $\protect\chi_1$ (solid) and imaginary part $\protect\chi %
_2$ (dashed) of the susceptibility $\protect\chi$ vs the probe light
detuning $\Delta_{p}$ in normalized units of $\Gamma _{A}$ according to: (a)
$\Delta =0$, $\Omega_1=1$ and $\Omega_2=0$; (b) $\Delta =0$, $\Omega_1=0$
and $\Omega_2=1$; (c) $\Delta =0$, $\Omega_1=\Omega_2=1$; (d) $\Delta =0$, $%
\Omega_1=\Omega_2=2$; (e,f) $\Delta =\pm 2$, $\Omega_1=\Omega_2=1$. The
Other parameters are given as: $\Gamma _{1}=\Gamma _{2}=10^{-4}$, $g\protect%
\sqrt{N}=100$.}
\label{figexp02}
\end{figure}

First, we consider the case that the two classical light fields have the
same detunings: $\Delta _{1}=\Delta _{2}\equiv \Delta $. Fig. \ref{figexp02}
shows $\chi _{1}$ and $\chi _{2}$ in such case versus the probe light
detuning 
$\Delta _{p}$ under $\Delta =0,\pm 2$ and different Rabi frequencies $\Omega
_{1,2}$ with the other parameters being fixed as $\Gamma _{1}=\Gamma
_{2}=10^{-4}$, $g\sqrt{N}=100$ (all in normalized units of $\Gamma _{A}$).
Seen from Figs. \ref{figexp02}, when $\Delta _{p}\rightarrow \Delta $, both $%
\chi _{1}$ and $\chi _{2}$ are almost equal to zero. This fact means that
the medium indeed becomes transparent when driven by the two classical
control fields as long as the system is prepared in the 3-photon resonance ($%
\Delta _{p}=\Delta _{1}=\Delta _{2}$) without the condition of exact
one-photon resonance: $\Delta _{i}\equiv 0$ ($i=p,1,2$). We also notice that
the width of the induced transparency window (which is determined by $\chi
_{2}$ in the near domain of $\Delta _{p}=\Delta $) also depends on the
concurrent interaction of Rabi frequencies $\Omega _{1,2}$. This is
intuitionistic and coincident with the previous work \cite%
{li-sun-prar,Scullybook}: under the case of 3-photon resonance, since each
classical control field as well as the relative level induces a transparency
window for the quantum probe field and appears the phenomenon of EIT
independently (under 2-photon resonance), such a system appears EIT
phenomenon too and results from the concurrent influence of these two
control fields.

\begin{figure}[h]
\begin{center}
\includegraphics[width=8cm,height=5.5cm]{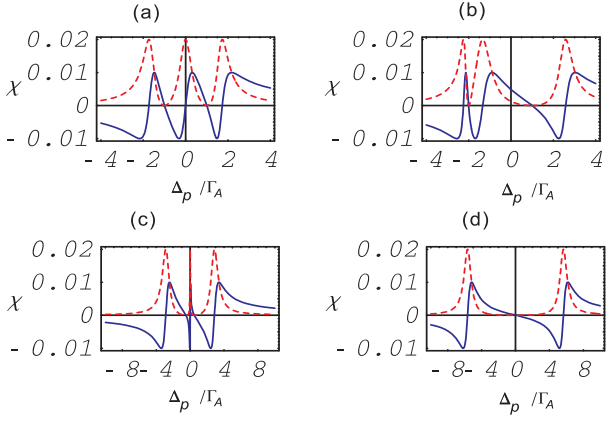}
\end{center}
\caption{Real part $\protect\chi_1$ (solid) and imaginary part $\protect\chi %
_2$ (dashed) of the susceptibility $\protect\chi$ vs the probe light
detuning $\Delta_{p}$ according to: (a) $\Delta_1 =1$, $\Delta_2 =-1$, $%
\Omega_1=\Omega_2=1$; (b) $\Delta_1 =1$, $\Delta_2 =-2$, $\Omega_1=2$, $%
\Omega_2=1/2$; (c) $\Delta_1 =0.5$, $\Delta_2 =-0.5$, $\Omega_1=2$, $%
\Omega_2=2$; (d) $\Delta_1 =0.05$, $\Delta_2 =-0.05$, $\Omega_1=4$, $%
\Omega_2=4$. The Other parameters are given as: $\Gamma _{1}=\Gamma
_{2}=10^{-4}$, $g\protect\sqrt{N}=100$.}
\label{figexp03}
\end{figure}

Then, let us consider the case of the two classical control light fields
having the different detunings: $\Delta _{1}\neq \Delta _{2}$. Fig. \ref%
{figexp03} shows the dependance of $\chi _{1}$ and $\chi _{2}$ on the probe
light detuning $\Delta _{p}$ in this case. There are two transparent windows
for the probe light field this time. These two windows appear near the
points of $\Delta _{p}=\Delta _{1}$ and $\Delta _{p}=\Delta _{2}$, with the
width of windows depending on $\Omega _{1}$ and $\Omega _{2}$ respectively.
When $\Delta _{1}\rightarrow \Delta _{2}$ and the Rabi frequencies are
strong enough, two transparency windows will overlap (see Fig. \ref{figexp03}%
(c)) or even become one (see Fig. \ref{figexp03}(d)). This results from the
fact that: when the quantum probe field together with one of the classical
control fields satisfies the 2-photon resonance condition, a transparency
window appears for the quantum probe field and so does the phenomenon of
EIT, since at the same time the effect of the other control field (which
does not satisfy the 2-photon resonance together with the probe field) is
small and can be ignored. In the next section, we will continue to calculate
the group velocity under this case and the case of 3-photon resonance.

%

%

\section{The Group Velocity of quantum probe light field}

Next we consider the property of refraction and absorption of the
single-mode probe light within the atomic ensemble medium in more detail. To
this aim we will analyze the complex refractive index
\begin{equation}
n(\omega )=\sqrt{\epsilon (\omega )}=\sqrt{1+\chi }\equiv n_{1}+in_{2}.
\label{n}
\end{equation}%
Where the real part $n_{1}$ of $n$ represents the refractive index of the
medium and the imaginary $n_{2}$ is the associated absorption coefficient.
Together with the formulae for the group velocity of the probe light
\begin{equation}
v_{g}(\Delta _{p})=\frac{c}{\mathrm{Re}[n+\omega \frac{\mathrm{d}n}{\mathrm{d%
}\omega }]}=\frac{c}{n_{1}+\omega \frac{\mathrm{d}n_{1}}{\mathrm{d}\omega }}
\label{vg1}
\end{equation}%
(where $c$ is the light velocity in vacuum) depending on the frequency
dispersion, one can obtain the explicit expression for the group velocity $%
v_{g}$ from Eqs. (\ref{chi-f}-\ref{vg1}) for arbitrary reasonable values of $%
\Delta _{p}$ and other parameters. Now, we consider the group velocity of
the probe light $v_{g}$ under the case of EIT. At this time the values of $%
\chi _{1}$ and $\chi _{2}$ are almost zero, and we obtain approximately
\begin{equation*}
n_{1}\simeq 1+\chi _{1}/2\rightarrow 1,\ \ n_{2}\simeq \chi _{2}\rightarrow
0,
\end{equation*}%
and the group velocity of probe light is given briefly as:
\begin{equation}
v_{g}(\Delta _{p})=\frac{c}{1-\frac{\omega }{2}\frac{\mathrm{d}\chi _{1}}{%
\mathrm{d}\Delta _{p}}}.  \label{vg2}
\end{equation}%
It is worth stressing that the above Eq. (\ref{vg2}) is effective only under
the case of EIT.

%
\begin{figure}[h]
\begin{center}
\includegraphics[width=8cm,height=3.0cm]{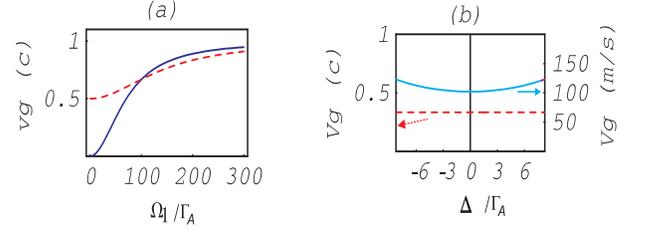}
\end{center}
\caption{The probe light group velocity $v_{g}$ under the case of 3-photon
resonance vs: (a) the Rabi frequency $\Omega_{1}$ [in normalized units] for $%
\Delta =0$ and $\Omega_{2}$ being assumed to vary synchronously with $%
\Omega_{1}$ (solid line) or $\Omega_{2}$ being given as $100$ (dashed line);
(b) the detuning $\Delta $ ($\equiv \Delta_p $) for $\Omega_1=\Omega_2=0.04$
(solid line), or $\Omega_1=\Omega_2=50$ (dashed line). The other parameters
are given as: $\Gamma _{1}=\Gamma _{2}=10^{-4}$ and $g\protect\sqrt{N}=100$.}
\label{figexp04}
\end{figure}

%
\begin{figure}[h]
\begin{center}
\includegraphics[width=5cm,height=3cm]{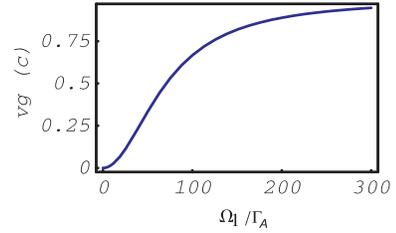}
\end{center}
\caption{The probe light group velocity $v_{g}$ (under EIT of non-3-photon
resonance: $\Delta _1\neq \Delta _2)$ vs the Rabi frequency $\Omega_{1}$ ($%
\equiv\Omega_{2}$) for $\Delta _p=\Delta _1=2$, $\Delta _2=-2$ and the other
parameters given as $\Gamma _{1}=\Gamma _{2}=10^{-4}$ and $g\protect\sqrt{N}%
=100$.}
\label{figexp05}
\end{figure}

According to Eq. (\ref{vg2}), the group velocity $v_{g}$ of the probe light
within the 3-photon-resonance atomic ensemble is shown in Fig. \ref{figexp04}%
. In Fig. \ref{figexp04}(a), the solid line shows the dependance of $v_{g}$
on Rabi frequencies $\Omega _{1,2}$ (assume $\Omega _{1,2}$ vary at the same
time), the dashed one shows the dependance of $v_{g}$ on Rabi frequencies $%
\Omega _{1}$ ($\Omega _{2}$ is fixed as $100$), with the other parameters
being given as $\Delta _{p}\equiv \Delta =0$, $\Gamma _{1}=\Gamma
_{2}=10^{-4}$, and $g\sqrt{N}=100$. Fig. \ref{figexp04}(b) shows the
dependance of $v_{g}$ on detuning $\Delta $ when the other parameters are
given as $\Delta _{p}\equiv \Delta $, $\Gamma _{1}=\Gamma _{2}=10^{-4}$, $g%
\sqrt{N}=100$, $\Omega _{1}=\Omega _{2}=50$ (dashed line) or $\Omega
_{1}=\Omega _{2}=0.04$ (solid line). This implies that: when $\Omega _{1}$
\textit{or} $\Omega _{2}$ is big (compared with $g\sqrt{N}$), $v_{g}$ is
relatively fast and insensitive to the common detuning $\Delta _{p}$ ($%
\equiv \Delta $) (see the dashed line in Fig. \ref{figexp04}); however when
both $\Omega _{1}$ \textit{and} $\Omega _{2}$ are small, $v_{g}$ is
relatively slow and sensitive to the common detuning $\Delta _{p}$ ($\equiv
\Delta $) (see the solid line in Fig. \ref{figexp04}).

We have also calculated $v_{g}$ under the 2-photon resonance (but not
3-photon resonance) EIT. \ Fig. \ref{figexp05} shows the dependance of $%
v_{g} $ on Rabi frequencies ($\Omega _{2}\equiv \Omega _{1}$ at this time),
when $\Delta _{p}\equiv \Delta _{1}=2$ $\neq \Delta _{1}=-2$ with the other
parameters being given as $\Delta _{p}\equiv \Delta =0$, $\Gamma _{1}=\Gamma
_{2}=10^{-4}$, and $g\sqrt{N}=100$. The result shows\ under this case, the
probe field group velocity $v_{g}$ also can been varied in the scope of ($%
0,c $) with $\Omega _{1,2}$ being varied.

This fact, the probe field group velocity $v_{g}$ decreases dramatically
with small $\Omega _{1,2}$, ensures that the technique as shown in Ref. \cite%
{li-zhang} is effective to accomplish the storage and retrieve of the probe
pulse. The storage process of such a technique is that: initially when the
probe field enters into the 3-photon-resonance atomic medium, the Rabi
frequency $\Omega $ is very large (relative to $g\sqrt{N}$) and $%
v_{g}\rightarrow c$; when $\Omega _{1,2}$ are reduced adiabatically to zero,
$v_{g}$ reduces to zero accordingly and then one can store the pulse in the
medium. Conversely, the retrieve process is the inverse one. That is, if one
wants to retrieve the probe pulse, he only needs to increase $\Omega $
adiabatically so as to increase $v_{g}$.

\section{Conclusion}

In this work, based on the dynamical algebra method of atomic collective
excitation  \cite{Sun-prl,li-sun-prar}, we have studied theoretically the
susceptibility and group velocity of a quantum probe light in a "3+1"-level
atomic ensemble under EIT. Our results show the quantum probe light group
velocity can been reduced dramatically under tiny values of $\Omega _{1,2}$.
This is very useful during the storage of the quantum probe light in such a
"3+1"-level atomic ensemble. Moreover, our results show that two
transparency windows for the probe light can occur in the case of two
classical control light fields having the different detuings to the relative
atomic transitions. Of course, in the practical experiment for store a
quantum light, the influence of atomic spatial motion or atomic collisions,
and the effects of buffer gases, should be taken into account. In the
present work, all of these effects are ignored as the perturbations for we
assume that the atomic ensemble is prepared under enough low temperature.

\textit{We acknowledge the support of the CNSF (grant No. 90203018), the
Knowledge Innovation Program (KIP) of the Chinese Academy of Sciences and
the National Fundamental Research Program of China (No. 2001CB309310). Y. L.
also thanks the support of National Natural Science Foundation of China with
No. 10447133.}

\end{document}